\pdfoutput=1

\documentclass[conference]{IEEEtran}
\ifCLASSINFOpdf
\else
\fi
\usepackage{xcolor}
\usepackage{cite}
\usepackage{amsmath}
\usepackage{amsfonts}
\usepackage[short]{optidef}
\usepackage{subfig}
\usepackage[pdftex]{graphicx}
\usepackage{autobreak}
\usepackage{xcolor}

\allowdisplaybreaks
\newcommand{\T}{\mathcal{T}}
\newcommand{\C}{\mathcal{C}}

\usepackage[top=0.7in, left=0.7in,right=0.7in, bottom=0.95in]{geometry}
\usepackage{comment}

\allowdisplaybreaks[4]

\begin{document}
%
\title{Shaping Next-Generation RAN Topologies to Meet Future Traffic Demands: A Peak Throughput Study}

\author{\IEEEauthorblockN{Paolo Fiore$^*$, Ilario Filippini$^*$ and Danilo De Donno$^\mathsection$}
\vspace{-0.2cm}
\IEEEauthorblockA{
\hfill\\
$^*$ANTLab - Advanced Network Technologies Laboratory, Politecnico di Milano, Milan, Italy \\
$^\mathsection$Milan Research Center, Huawei Technologies Italia S.r.l, Milan, Italy\\
Email: $\{$paolo.fiore, ilario.filippini$\}$@polimi.it, danilo.dedonno@huawei.com
}
\vspace{-10mm}
}


\maketitle

\begin{abstract}
Millimeter-Wave (mm-Wave) Radio Access Networks (RANs) are a promising solution to tackle the overcrowding of the sub-6 GHz spectrum, offering wider and underutilized bands. However, they are characterized by inherent technical challenges, such as a limited propagation range and blockage losses caused by obstacles. Integrated Access and Backhaul (IAB) and Reconfigurable Intelligent Surfaces (RIS) are two technologies devised to face these challenges. This work analyzes the optimal network layout of RANs equipped with IAB and RIS in real urban scenarios using MILP formulations to derive practical design guidelines. 
In particular, it shows how optimizing the peak user throughput of such networks improves the achievable peak throughput, compared to the traditional mean-throughput maximization approaches, without actually sacrificing mean throughputs. In addition, it indicates star-like topologies as the best network layout to achieve the highest peak throughputs.
\end{abstract}
\IEEEpeerreviewmaketitle

\vspace{-1.5mm}
\section{Introduction}

Since its introduction in 3GPP Release 15, the millimeter-Wave (mm-Wave) radio spectrum 
has been sought over as a launchpad to reach previously unachievable transmission rates for end users. With the prospect of bandwidths orders of magnitude larger than the ones possible in LTE and 5G sub-6 GHz
, combined with a sparsely used spectrum, this frequency range has become very attractive to operators, vendors, and researchers to answer the pressing issue of spectrum overcrowding in sub-6 GHz and to give traction to innovation and development in mobile radio communications~\cite{rapparport2014,Akdeniz2014}.\\
Moreover, the need for a more suitable frequency range for high-speed/low-latency use cases spawns from numerous forecasts, which consistently predict an ever-increasing volume of traffic consumed by mobile devices. As reported in~\cite{ericssonReportNov2022}, it is expected that, by 2028, the total mobile traffic will increase three-fold compared to 2023, and all of the data traffic growth will come from 5G New Radio (NR) connections, leaving far behind the usage of previous generations of mobile Radio Access Networks (RAN), such as GSM (2G), UMTS (3G), and LTE (4G).

The public expectation on mm-Wave frequencies must face the inherent challenges of transmitting with limited propagation range due to severe path loss and blockage loss caused by obstacles at such a high frequency~\cite{8254900}. Suppose an obstacle interrupts a mm-Wave radio link. In that case, the high reflectivity of materials found in urban areas causes the wave to be deflected in unwanted directions, reducing the probability of being detected by the designated receiver~\cite{rapparport2014}.\\
One way to partially solve the propagation shortcomings of mm-Wave frequencies consists in densifying the Radio Access Network (RAN) by installing a greater number of base stations in the considered area. This solution comes with increased installation costs proportional to the desired level of densification. Integrated Access and Backhaul (IAB), a paradigm standardized yet in Release 16, can mitigate the limitations of this approach. Backhaul links between base stations are relocated from the more expensive underground fiber cabling to the radio spectrum. In this way, all radio links can be shortened by deploying simpler and cheaper devices than full-fledged base stations. This process effectively reduces path loss and, at the same time, can be less taxing on the capital expenditure (CAPEX), saving up to 85$\%$ of installation costs as there is no need for wired connections and trenching~\cite{10.3389/frcmn.2021.647284,938713,8514996,9040265}. Nevertheless, massive RAN densification still represents a challenging topic spawning the search for other alternative solutions.

The recent topic of metasurfaces, particularly Reconfigurable Intelligent Surfaces (RIS), is gaining momentum in the mm-Wave research and industry community~\cite{9720231} to boost throughput~\cite{Moro2021} and resilience to obstacle outages~\cite{9771934,9930587}. RIS are planar surfaces made of small radiating elements and have been proven cheaply mass-producible. RIS can steer the impinging waves to any direction in their Field of View (FoV) in a quasi-passive way (also addressing power consumption concerns), and can exploit alternative radio paths that, as previously mentioned, are mostly unavailable, managing to turn around static obstacles (e.g., buildings) and limiting the impact of sudden obstacles (e.g., vehicles and pedestrians).\\
The technologies above are promising enough to have been an object of study in recent publications~\cite{tsilipakos2020toward,di2019smart}, and a detailed analysis is warranted to evaluate their impact on the connectivity service from a networking perspective. This has produced Mixed-Integer Linear Programming (MILP) models, whose solutions mathematically describe how and where each device should be installed to deploy a RAN that maximizes a preselected indicator. This allows to unequivocally measure the network performance and assess the contribution of these new technologies. Traditionally, the mean user throughput has been employed as the main parameter for network optimization~\cite{6678102}.

However, in~\cite{sandvineReport2022,sandvineReport2023}, it has been recently stated that mobile data traffic of type \textbf{video} has increased from $53.72$ (2021)~\cite{sandvineReport2022} to $67.60\%$ (2022)~\cite{sandvineReport2023} of the total mobile traffic and is set to keep increasing in the following years. Therefore, it is time to evaluate different video traffic-friendly metrics as the focus of optimization, even more so considering the different propagation conditions of mm-Waves compared to commonly used lower frequencies.\\
With such an abundance of bandwidth available in the mm-Wave spectrum, ultra-high throughput transmissions will transfer large data volumes in very-short time periods, which will be then consumed in the idle period before the next large-volume burst. A shorter occupation of the RAN resources leads more users to transmit and receive at the maximum rate available. Thus, peak user throughput becomes a potentially crucial optimization parameter. In addition, speed tests are now an extremely common tool to evaluate the quality of a mobile operator; hence, maximizing the peak user throughput will be very effective.

This work, to the best of our knowledge, is the first to analyze how an urban RAN topology that is planned by optimizing the peak user throughput compares against a more traditional mean throughput approach.
Under the optimality guarantees of MILP models, we show that deploying a peak-throughput optimal topology yields mean-throughput results similar to traditional methods while significantly enhancing the maximum achievable throughput, all at an equal installation cost.\\
The remainder of this paper is structured as follows: Section~\ref{sec:system_model} focuses on the detailed description of the components of the considered network scenario, and Section~\ref{sec:milp_models} describes the MILP network planning models used; Section~\ref{sec:results} compares the results obtained by the conventional approach and our proposal; final remarks in Section~\ref{sec:conclusion} conclude the paper.
\label{sec:intro}

\vspace{-1.85mm}

\section{System Model} 
\begin{figure}
    \centering
    \includegraphics[width=0.95\columnwidth]{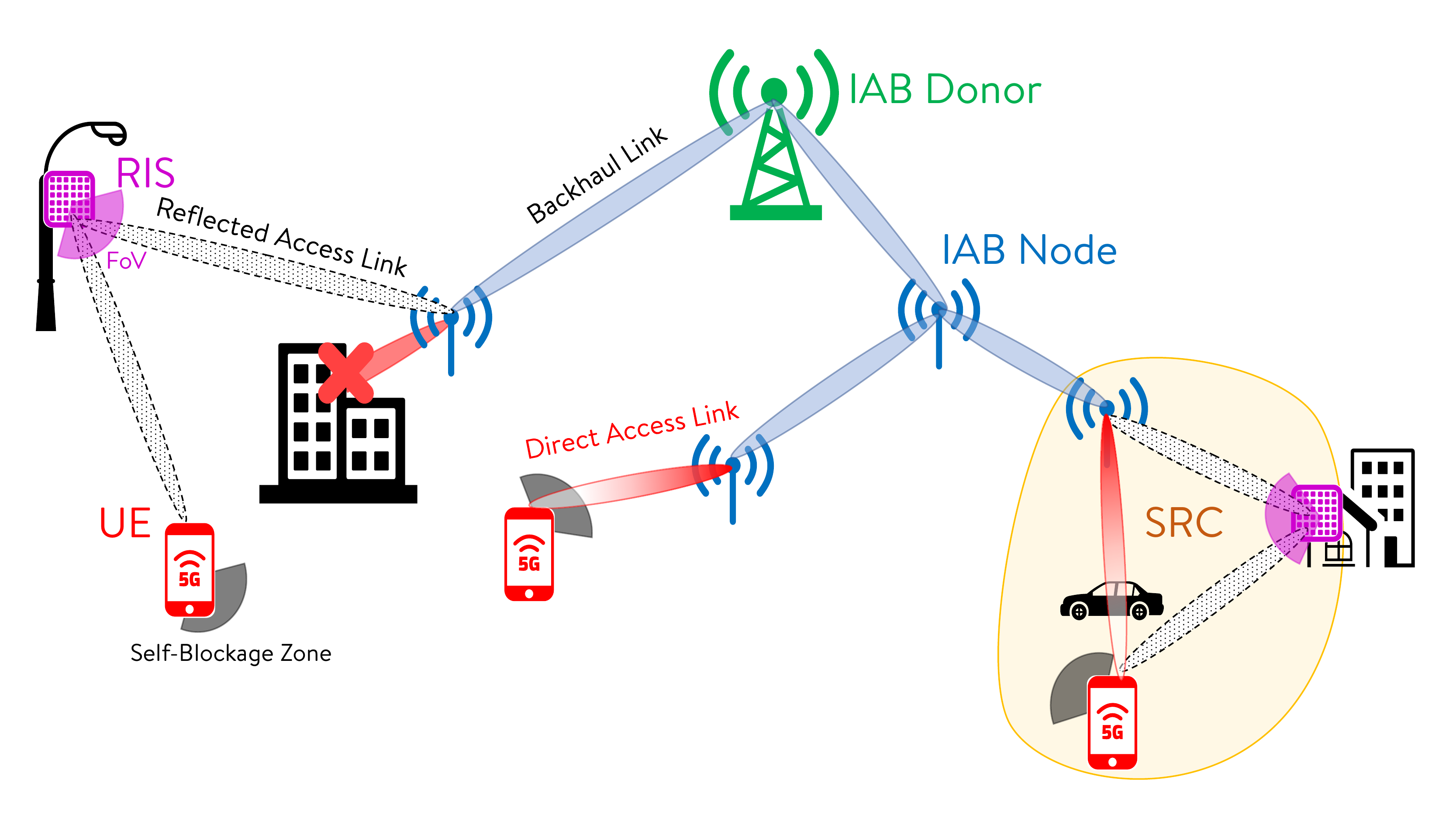}
    \vspace{-3mm}
    \caption{\small RAN scenario with different types of obstacles.}
    \label{fig:scenario}
    \vspace{-6mm}
\end{figure}
In this section, we provide a description of a RIS-empowered mm-Wave IAB Radio Access Network, shown in Figure~\ref{fig:scenario}. We describe in detail the main elements of such networks, their behavior, and how they relate to each other in the system model.

An IAB network consists of: User Equipments (UE) that must be served; a single IAB Donor, a full-fledged Base Station (BS) and the only node in the RAN cabled to the rest of the network, and IAB Nodes, simple BS capable of giving access to UEs and wireless relaying data from/to other IAB Nodes. Backhaul links between IAB Nodes and access links serving the UEs operate at the same frequency range (\textit{in-band backhauling}). An IAB network assumes a tree topology, as in 3GPP specifications~\cite{3GPPTR38.874}, enabling end-to-end connections between the Donor and the UEs in a multi-hop fashion. Due to space limitations, only downlink traffic flow has been considered in this work. Nevertheless, simple amendments can be added to straightforwardly extend it to consider uplink traffic.

All links in the RAN use advanced beamforming to improve propagation at the mm-Wave range: thus, given the narrow nature of the beams, in concert with a half-duplex operation mode of every device and a continuous Time Division Multiplexing (TDM) approach\footnote{We considered the timesharing resources to be allocated in a continuous temporal frame instead of a real-scenario discrete resource block}, the impact of mutual interference is typically minimal. Therefore, we consider interference between different links negligible~\cite{devoti2020,7342886}.

The use of RISs in an IAB Network is mainly motivated by their ability to impact EM propagation in order to improve channel conditions in case of obstacles. RISs operate as passive beamformers, capable of redirecting an incident radio signal toward a desired direction. UEs can be served by a single direct link from an IAB Node or by a Smart Radio Connection (SRC), a triplet involving a UE and both an IAB Node and a RIS. IAB Nodes are expected to dynamically command and change SRC configurations~\cite{9720231} to activate the reflection through the RIS, within their FoV, to improve throughput during obstacle obstruction, and thus resilience to obstacle blockage.

We adopt the mm-Wave channel model provided in~\cite{https://doi.org/10.48550/arxiv.2211.08033}, which includes the effect of RISs, as well as the physical characterization of the involved devices in the RAN (IAB Donor, IAB Node, RIS, and UE). We consider nomadic obstacles modeled as in ~\cite{https://doi.org/10.48550/arxiv.2211.08033}
, where the blockage probability and blockage loss are calculated from Monte-Carlo simulations and then fitted to derive a probability distribution.
We also consider the self-blockage zone, modeled as in 3GPP specifications, which consists of a circular sector centered in the UE position where the signal attenuation is increased to model the effect of users' body~\cite{GRANW2018study}. 

We consider static and nomadic obstacles. Static obstacles (i.e., buildings) can be avoided during the network planning phase so that IAB Nodes can be connected only if in Line-of-Sight (LoS) conditions. Vice versa, the heights of nomadic obstacles are typically smaller than those of IAB Nodes' and RIS' installation sites, therefore their presence will affect only access links from the access nodes (IAB Nodes and RISs) to the UEs. For these reasons, we assume all backhaul links are always in Line-of-Sight (LoS) conditions, and their capacity (in Mb/s) is constant. On the other hand, access links can be interrupted by nomadic obstacles, the user's body, or both; therefore, their average capacity is weighted by the probability of being in every possible blockage state (including both direct link and SRC links). Moreover, when an SRC is available, only the best link between the direct IAB-UE and the reflected RIS-UE is selected as the access link.\\
\label{sec:system_model}

\vspace{-0.3em}
\section{MILP Models}
\label{sec:milp_models}
We now detail two different MILP formulations to provide optimal IAB network topology: the Mean-Throughput Formulation, a baseline formulation in which the objective function maximizes the mean user throughput, and the Peak-Throughput Formulation, an extension of the baseline, in which the peak user throughput becomes the objective of the maximization. Both formulations share a common notation, defined as follows.

Adopting the standard approach used in the literature\cite{doi:https://doi.org/10.1002/9781118511305.ch7}, we define a set $\C$ of Candidate Sites (CS) where a network device (the IAB Donor, an IAB Node, or a RIS) can be installed over an urban area planned to be covered by a mm-Wave network. Inside $\C$, two CSs $\hat{c}$ and $\Tilde{c}$ are selected for the installation of two fixed special devices: $\hat{c}$ is always reserved for the IAB Donor. In contrast, $\Tilde{c}$ is a placeholder CS for a "fake" RIS. This method lets the solver decide whether to assign a real SRC (a BS, a UE, and a RIS) or a "fake" SRC (a BS, a UE, and the "fake" RIS), thus a direct connection, to each UE, without the need to introduce additional variables to characterize the two different kinds of access connections. Test Points (TP), centroids of traffic mimicking the geographical distribution of UEs in the area, are represented by set $\T$.

All physical characteristics of an SRC, such as SNR and blockage loss, are encapsulated in the binary activation parameters $\Delta$ and the achievable rate parameters $C \in \mathbb{R^+}$. Parameter $\Delta_{t,c,r}^{SRC}$ equals 1 when an SRC can be established between TP $t \in \T$, an IAB device installed in $c \in \C$, and a RIS in CS $r \in \C$. Similarly, parameter $\Delta_{c,d}^{BH}$ indicates the availability of the inband-backhaul link between IAB nodes $c,d \in \C$. When an SRC can be established, $C_{t,c,r}^{SRC}$ is the weighted average of capacities calculated in various states of blockage, while
$C_{t,c,r}^{RIS} \in \mathbb{R^+}$ is the achievable rate of SRC ($t,c,r$) when only the reflected path through the RIS is available.
Finally, $C_{c,d}^{BH} \in \mathbb{R^+}$ defines the capacity of a backhaul link between two nodes $c,d \in \C$. 
A minimum amount of demand $D$, measured in Mb/s, must be served to each TP.

Every RIS has an associated azimuthal angle $F$, representing its Field of View (FoV), where both an IAB Node and a TP must lie in to be able to use the RIS\footnote{The elevation angle of the FoV of the RIS is managed in pre-processing given the fixed height of the involved devices in the planning scenario.}. To evaluate if two devices fall within the FoV of the RIS, parameters $\Phi_{r,t}^{\text{A}},\Phi_{r,c}^{\text{B}} \in \left[0,2\pi\right]$ must be defined: the former represents the angle between RIS $r \in \C$ and TP $t \in \T$, and the latter the angle between RIS $r$ and BS $c$, with $r,c \in \C$.

Finally, the selection of network devices to deploy is constrained by budget $B$, and $P^{\text{IAB}}$ and $P^{\text{RIS}}$ are the prices for IAB Nodes and RISs, respectively.{{\begin{table}[htbp]
\caption{Decision variables for MTF and PTF}
\begin{center}
\begin{tabular}{|c|c|}
\hline
\textbf{Variable}&\textbf{Description} \\ 
\hline
\text{$y_c^{\text{DON}}\in\{0,1\}$}&\text{Installation of IAB Donor in CS $c \in \C$.} \\
\hline
\text{$y_c^{\text{IAB}}\in\{0,1\}$}&\text{Installation of IAB Node in CS $c \in \C$.} \\
\hline
\text{$y_c^{\text{RIS}}\in\{0,1\}$}&\text{Installation of RIS in CS $c \in \C$.} \\
\hline
\text{$x_{t,c,r}\in\{0,1\}$}&\text{Activation of SRC ($t,c,r$), $t\in\T, c,r\in\C$.} \\
\hline
\text{$z_{c,d}\in\{0,1\}$}&\text{Activation of backhaul link ($c,d$), $c,d\in\C$.} \\
\hline
\text{$f_{c,d}\in\mathbb{R^+}$}&\text{Traffic on backhaul link ($c,d$), $c,d\in\C$.} \\
\hline
\text{$g_{t,c,r}\in\mathbb{R^+}$}&\text{Traffic on SRC ($t,c,r$), $t\in\T, c,r\in\C$.} \\
\hline
\text{$w_c\in\mathbb{R^+}$}&\text{Total traffic to the IAB Donor, $c\in\C$.} \\
\hline
\text{$t_c^{\text{TX}}\in\left[0,1\right]$}&\text{Transmission time ratio of BS installed in CS $c\in\C$.} \\
\hline
\text{$t_c^{\text{RX}}\in\left[0,1\right]$}&\text{Reception time ratio of BS installed in CS $c\in\C$.} \\
\hline
\text{$\phi_c\in\left[0,2\pi\right]$}&\text{Azimuthal orientation of RIS installed in CS $c\in\C$.} \\
\hline
\multicolumn{2}{|c|}{\textbf{PTF flow variables}}\\
\hline
\text{$f_{t,c,d}^\text{X}\in\mathbb{R^+}$}&\text{Extra traffic on backhaul link ($c,d$), $c,d\in\C$.} \\
\hline
\text{$g_{t,c,r}^\text{X}\in\mathbb{R^+}$}&\text{Extra traffic on SRC ($t,c,r$), $t\in\T, c,r\in\C$.} \\
\hline
\text{$w_{t,c}^\text{X}\in\mathbb{R^+}$}&\text{Total extra traffic to the IAB Donor, $c\in\C$.} \\
\hline
\end{tabular}
\label{tab:vars}
\end{center}
\end{table}}}The network planning model's solution consists in assigning a value to the decision variables in Table~\ref{tab:vars}, as per the goal stated by the selected objective function. Said variables determine which devices will be installed and where ($y_c^{\text{DON}}, y_c^{\text{IAB}}, y_c^{\text{RIS}}$), how they will be connected to each other ($x_{t,c,r}, z_{c,d}$), how much traffic will flow through each link ($ f_{c,d}, g_{t,c,r}, w_c$), how the time resources of each BS are employed ($t_c^{\text{TX}}, t_c^{\text{RX}}$), and how the RIS will be oriented ($\phi_c$).

\subsection{Mean-Throughput Formulation (MTF)}\label{sub:mconst}
The MTF is defined by the following objective and constraints:
{
{
\small
\begin{subequations}
\begin{equation}
\max \sum_{t \in \T, c,r \in \C} g_{t,c,r}\label{opt:obj}\\
\end{equation}
\vspace{-0.7cm}
\begin{flalign}
&\text{s.t.:}&\nonumber\\
&y_c^\text{IAB} + y_c^\text{RIS}\leq 1, & \forall c \in \C,\label{opt:ris_iab_act}\\
&y_c^\text{DON} \leq y_c^\text{IAB},& \forall c \in \C,\label{opt:don_act}\\
&\sum_{c \in \C}y_c^\text{DON}\leq 1,&\label{opt:only_1_don}\\
& y_{\Tilde{c}}^{\text{RIS}}  \geq 1,&\label{opt:fake_ris}\\
& y_{\hat{c}}^{\text{DON}}  \geq 1,&\label{opt:fixed_donor}\\
&z_{c,d} \leq \Delta^\text{BH}_{c,d}\left( y_c^\text{IAB} + y_d^\text{IAB} \right)/2,&\forall c,d \in \C,\label{opt:bh_link_act}\\
&x_{t,c,r} \leq \Delta^\text{SRC}_{t,c,r}\left( y_c^\text{IAB} + y_r^\text{RIS} \right)/2,& \forall t \in \T, c,r \in \C,\label{opt:src_act}\\
&\sum_{c,r \in C}x_{t,c,r} = 1,&\forall t \in \T,\label{opt:one_src}\\
&\sum_{d \in \C}z_{d,c} \leq 1-y_c^\text{DON},&\forall c \in \C,\label{opt:spanning_tree}\\
&\sum_{c \in \C \setminus \{\hat{c},\Tilde{c}\}}\left( P^\text{IAB}y_c^\text{IAB} + P^\text{RIS}y_c^\text{RIS}\right) \leq B,&\label{opt:budget}
\end{flalign}
\vspace{-0.5cm}
\begin{flalign}
&w_c + \sum_{d \in \C}\left( f_{d,c} - f_{c,d} \right) - \sum_{\substack{t \in \T\\ r \in \C}}g_{t,c,r} = 0,&\forall c \in \C,\label{opt:flow_balance}
\end{flalign}
\vspace{-0.7cm}
\begin{flalign}
&f_{c,d} \leq C_{c,d}^\text{BH}z_{c,d}, &\forall c,d \in \C,\label{opt:flow_act}\\
& D x_{t,c,r} \leq g_{t,c,r} \leq C_{t,c,r}^\text{SRC} x_{t,c,r}, &\forall t \in \T, c, r \in \C,\label{opt:src_cap}\\
& w_c \leq M^{\text{MAX}} y_c^{\text{DON}}, &\forall c \in \C,\label{opt:core_cap}
\end{flalign}
\vspace{-0.7cm}
\begin{flalign}
\footnotesize
&t_c^\text{TX} = \sum_{d \in \C} \frac{f_{c,d}}{C_{c,d}^\text{BH}} + \sum_{\substack{t \in \T\\r\in \C}} \frac{g_{t,c,r}}{C_{t,c,r}^\text{SRC}}, &\forall c \in \C,\label{opt:tx_time}\\
&t_c^\text{RX} = \sum_{d \in \C} \frac{f_{d,c}}{C_{d,c}^\text{BH}}, &\forall c \in \C,\label{opt:rx_time}
\end{flalign}
\vspace{-0.5cm}
\begin{flalign}
&t_c^\text{TX} + t_c^\text{RX}\leq y_c^\text{IAB}, &\forall c \in \C,\label{opt:tdm}\\
&\sum_{\substack{t\in \T\\ c \in \C}}\frac{g_{t,c,r}}{C_{t,c,r}^\text{RIS}} \leq y_r^\text{RIS},&\forall r \in\C\setminus {\Tilde{c}},\label{opt:ris_tdm}\\
&\phi_r\geq \Phi^{\text{A}}_{r,t} - F/2 - 2\pi(1- x_{t,c,r}),&\forall t \in \T, c,r \in \C\setminus {\Tilde{c}},\label{opt:or1}\\
&\phi_r\leq \Phi^{\text{A}}_{r,t} + F/2 + 2\pi(1- x_{t,c,r}),&\forall t \in \T, c,r \in \C\setminus {\Tilde{c}},\label{opt:or2}\\
&\phi_r\geq \Phi^{\text{B}}_{r,c} - F/2 - 2\pi(1- x_{t,c,r}),&\forall t \in \T, c,r \in \C\setminus {\Tilde{c}},\label{opt:or3}\\
 &\phi_r\leq \Phi^{\text{B}}_{r,c} + F/2 + 2\pi(1- x_{t,c,r}),&\forall t \in \T, c,r \in \C\setminus {\Tilde{c}}.\label{opt:or4}
\end{flalign}
\end{subequations}
}

}
The MTF objective function (\ref{opt:obj}) maximizes the sum-throughput of all UEs\footnote{The value of the objective function is divided by $|\T|$ in post-processing to obtain the mean user throughput.}.

\textit{Deployment constraints (\ref{opt:ris_iab_act}-\ref{opt:fixed_donor}):} Constr.~(\ref{opt:ris_iab_act}) guarantees mutual exclusivity between IAB Nodes and RISs in a specific CS $c \in \C$, constr.~(\ref{opt:don_act}) enables the possibility of an IAB Node to be promoted to a Donor, constr.~(\ref{opt:only_1_don}) guarantees at most a single Donor, while constr.~(\ref{opt:fake_ris}) and constr.~(\ref{opt:fixed_donor}) install the "fake" RIS and the Donor in $\Tilde{c},\hat{c} \in \C$, respectively.

\textit{Link-Activation constraints (\ref{opt:bh_link_act}-\ref{opt:spanning_tree}):} Constr. (\ref{opt:bh_link_act}) activates a backhaul link between two CSs $c,d \in \C$ if both of them have an IAB Node installed and the physical characteristics of the potential link are favorable ($\Delta_{c,d}^{BH} = 1 $). In the same way, in (\ref{opt:src_act}), an SRC is activated if a BS is installed in $c$ and a RIS in $r$, and binary parameter $\Delta_{t,c,r}^{SRC}$ is equal to 1, $c,r \in \C$. (\ref{opt:one_src}) forces each TP to be served by a single SRC, and (\ref{opt:spanning_tree}) guarantees the deployment of a tree topology.

\textit{Budget constraint:} Constr. (\ref{opt:budget}) limits the acquisition of devices to the budget $B$\footnote{This constraint makes sure not to consider in the budget the IAB Donor (which is a fixed cost and thus not included in the variable budget) and the "fake" RIS (which is not an actual device but rather a way to keep the formulation compact and easy to manage).}.

\textit{Flow constraints (\ref{opt:flow_balance}-\ref{opt:core_cap}):} Const. (\ref{opt:flow_balance}) guarantees flow balance at any BS in the tree, const. (\ref{opt:flow_act}) upper bounds backhaul link flows to link capacities, const. (\ref{opt:src_cap}) imposes both the minimum demand $D$ and the limit of maximum SRC capacity. Finally, constr. (\ref{opt:core_cap}) limit the traffic entering the RAN from the Core Network, under the assumption that this quantity cannot be more than the capacity of the best-performing link coming out of the Donor, indicated by $M^\text{MAX}$. Note that, although not strictly necessary, these types of constraints help reduce the solution time by tightly shaping the solution space.

\textit{Resource-sharing constraints (\ref{opt:tx_time}-\ref{opt:ris_tdm}):} Constr. (\ref{opt:tx_time}) defines the timeshare (assuming $1$ to be $100$\% of the available time) dedicated to transmission for any BS (in both the access and backhaul phases). Similarly, constr. (\ref{opt:rx_time}) defines the reception timeshare. Constr. (\ref{opt:tdm}) enforces half-duplex operation mode considering both transmission and reception timeshares, while constr. (\ref{opt:ris_tdm}) manages the timeshare of a RIS installed in CS $r\in\C$ among different SRCs $\{(t_1,c_1,r),\cdots,(t_n,c_n,r)\},t_1\cdots t_n\in\T,c_1\cdots c_n,r\in\C$.

\textit{RIS-orientation constr. (\ref{opt:or1}-\ref{opt:or4}):} These constraints set the value of the orientation variable $\phi_c$ dependent according to the angles between the involved devices and force angles of reflection links to lie within the FoV of the RIS, if any.\footnote{These constraints do not affect the status of the "fake" RIS, since it is not an actual device and does not need an orientation.}

\subsection{Peak-Throughput Formulation (PTF)}\label{sub:pform}
The Peak-Throughput Formulation extends the Mean-Throughput formulation by adding some variables and constraints. The other parts of the model are inherited from the previous formulation.

As mentioned in Section~\ref{sec:intro}, when users manage to establish a peak-throughput connection, the event is akin to a traffic burst, characterized by short duration and no correlation with other users' traffic. For this reason, the allocation of the extra traffic enabled by the peak-throughput is fundamentally different from the average traffic demand $g_{t,c,r}$, and must be modeled with a distinct approach. While being routed on different links from the IAB Donor to the final UE, the average demand $g_{t,c,r}$ for SRC ($t,c,r$) shares the same time resources with all the other SRCs in its path. The extra traffic $g_{t,c,r}^\text{X}$, on the other hand, does not share the resources with the other SRCs, but is allocated as if any considered UE is the only one in the network and can reserve all the capacity that is not used by the guaranteed traffic $g_{t,c,r}$.

The extra traffic of an SRC ($t,c,r$) is defined by the \textbf{spare capacities} of the links in its route from the IAB Donor to the UE; its value is the one of the BS with the least resources available (\textit{bottleneck BS}). The timeshares which are already reserved for the mean-throughput traffic of all UEs are still guaranteed by constraints (\ref{opt:tx_time}-\ref{opt:ris_tdm}) and thus remain untouched.

Therefore, we add further flow variables to capture peak-throughput traffic. They are listed in Table~\ref{tab:vars}. This model does not need to maximize the mean user throughput; all instances of variable $g_{t,c,r}$ in MTF (in constraints \ref{opt:flow_balance},\ref{opt:src_cap},\ref{opt:tx_time}, and \ref{opt:ris_tdm}) can be replaced with $D x_{t,c,r}$, resulting in a leaner formulation.\\
Objective function and constraints characterizing PTF are:
{
{
\small
\begin{subequations}
\begin{equation}
\max \sum_{t \in \T, c,r \in \C} g_{t,c,r}^\text{X}\label{opt_e:obj}\\
\end{equation}
\vspace{-0.7cm}
\begin{flalign}
&\text{s.t.:}&\nonumber\\
&w_{t,c}^\text{X} + \sum_{d \in \C}\left( f_{t,d,c}^\text{X} - f_{t,c,d}^\text{X} \right) - \sum_{\ r \in \C}g_{t,c,r}^\text{X} = 0,&\hspace{-0.45cm}\forall t \in \T, c \in \C,\label{opt_e:flow_balance}
\end{flalign}
\vspace{-0.4cm}
\begin{flalign}
&f_{t,c,d}^\text{X} \leq C_{c,d}^\text{BH}z_{c,d}, &\forall t \in \T, c,d \in \C,\label{opt_e:flow_act}\\
& g_{t,c,r}^\text{X} \leq C_{t,c,r}^\text{SRC} x_{t,c,r}, &\forall t \in \T, c, r \in \C,\label{opt_e:src_cap}\\
& w_{t,c}^\text{X} \leq M^{\text{MAX}} y_c^{\text{DON}}, &\forall t \in \T, c \in \C,\label{opt_e:core_cap}
\end{flalign}
\vspace{-0.5cm}
\begin{flalign}
\footnotesize
&\sum_{r \in \C}\frac{g_{t,c,r}^{\text{X}}}{C_{t,c,r}^{\text{SRC}}} + \sum_{d \in \C} \frac{f_{t,d,c}^{\text{X}}}{C_{d,c}^{\text{BH}}}\leq y_c^{\text{IAB}} - t_c^{\text{TX}} - t_c^{\text{RX}},&\hspace{-0.45cm}\forall c \in \C,t \in \T,\label{opt_e:tdm_e_acc}\\
&\sum_{d \in \C}(\frac{f_{t,d,c}^{\text{X}}}{C_{d,c}^{\text{BH}}} + \frac{f_{t,c,d}^{\text{X}}}{C_{c,d}^{\text{BH}}}) \leq y_c^{\text{IAB}} - t_c^{\text{TX}} - t_c^{\text{RX}}, &\hspace{-0.45cm}\forall c \in \C,t \in \T,\label{opt_e:tdm_e_bh}\\
&\sum_{c \in \C}\frac{g_{t,c,r}^{\text{X}}}{C_{t,c,r}^{\text{RIS}}}\leq y_{r}^{\text{RIS}} - \sum_{\substack{\tau \in \T\\ c \in \C}}\frac{D x_{\tau,c,r}}{C_{\tau,c,r}^{\text{RIS}}},& \hspace{-0.45cm}\forall r \in \C \setminus {\Tilde{c}}, t \in \T\label{opt_e:tdm_e_ris}.
\end{flalign}
\end{subequations}
}

}
The objective function (\ref{opt_e:obj}) maximizes the sum of the users' peak throughputs instead of the mean throughput of the previous formulation.
Constr. (\ref{opt_e:flow_balance}), similar to constr. (\ref{opt:flow_balance}), imposes the flow balance of the extra traffic routing through the tree; constr. (\ref{opt_e:flow_act}-\ref{opt_e:core_cap}) are the counterparts of capacity-related constr. (\ref{opt:flow_act}-\ref{opt:core_cap}), but involving the peak-throughput traffic. Finally, constr. (\ref{opt_e:tdm_e_acc}-\ref{opt_e:tdm_e_ris}) respectively model the BS' timeshare assigned to peak-throughput traffic for a BS involved in an SRC (\ref{opt_e:tdm_e_acc}), for a BS only involved in peak-throughput traffic backhauling (\ref{opt_e:tdm_e_bh}), and for a RIS (\ref{opt_e:tdm_e_ris}).
\subsection{Peak-Throughput Heuristics}\label{sub:heu}
The additional complexity introduced by peak-throughput traffic resulted in a formulation that was too hard to be solved in a reasonable amount of time (e.g., a single instance reached an unsatisfactory optimality gap of $30\%$ in one hour); therefore, we have developed a heuristic approach to speed up solving time while keeping a high level of quality of solutions. The technique consists in reducing the $M^\text{MAX}$ parameter, representing the highest-capacity backhaul link of the IAB Donor, to a fraction of its actual value, a sort of \textit{forced bottleneck} approach. This adjustment significantly reduces the solution space, speeding up the \textit{branch-and-cut} phase of the solution search.

This heuristic approach has been validated by comparing its results to those of the exact formulation in a set of smaller scenarios. Due to space limitations, we do not report the complete performance analysis. Nevertheless, the two approaches produced very similar performance trends; the heuristic approach reduces the solution time to an average of 6 seconds within a gap smaller than $5\%$.

\section{Results}
\label{sec:results}
In this section, we will compare the performances of the network scenarios optimized according to the mean-throughput formulation (MTF) against the ones obtained by applying the peak-throughput formulation (PTF), solved using the heuristic approach in Section~\ref{sub:heu}. We first investigate mean and peak throughputs achieved by both formulations, then we compare topological aspects to obtain a comprehensive performance analysis of the strengths and weaknesses of each model. The solutions are also iterated over progressively increasing values of the available budget to observe how much CAPEX will impact the quality of the solutions found.

The instances subject to network planning optimization are defined by a $150 m$ radius hexagonal cell deployment area centered around a randomly chosen point within a full 3D representation of the metropolitan area of Milan~\cite{MilanoGeoportale}, which includes buildings as opaque static obstacles of an actual urban scenario. The installation of the IAB Donor is fixed to the leftmost vertex of the hexagon; 25 CSs and 15 TPs are then randomly placed in the area according to surrounding buildings. The IAB Donor and the IAB Nodes are composed of three $120^{\circ}$ sectors with $16x12$ element panel arrays ($12x8$ for the IAB Node) and a $58 dBm$ EIRP ($51 dBm$ for the IAB Node), a carrier frequency of $28 GHz$ and a bandwidth of $200 MHz$. The UEs are modeled as a $2$x$2$ antenna array. RISs are made of $100$x$100$ passive elements in a rectangular array, with a FoV of $120^{\circ}$. All devices have $\frac{\lambda}{2}$ spacing in both directions between elements. The IAB Donor is set at a $25 m$ height, the IAB Nodes at $6 m$, the RISs at $3 m$, and the UEs at $1.5 m$. The price of IAB Nodes is normalized at $1$ unit of cost, while the expected inexpensive production process of RISs compared to IAB Nodes is encapsulated in a price ten times smaller at $0.1$. The total available budget spans from $6$ to $20$ units, with steps of $0.2$. Minimum UE demand is fixed at $150 Mbps$. All the link capacities are computed according to Shannon's capacity. All the results are averaged over $80$ random deployment instances generated through MATLAB and solved with IBM ILOG CPLEX. The maximum optimality gap between the feasible solutions found and their respective continuous relaxation upper bound is fixed at $5\%$.\\
\begin{figure}[!t]
\centering
\subfloat[Average installed devices]{\includegraphics[width=1.6in]{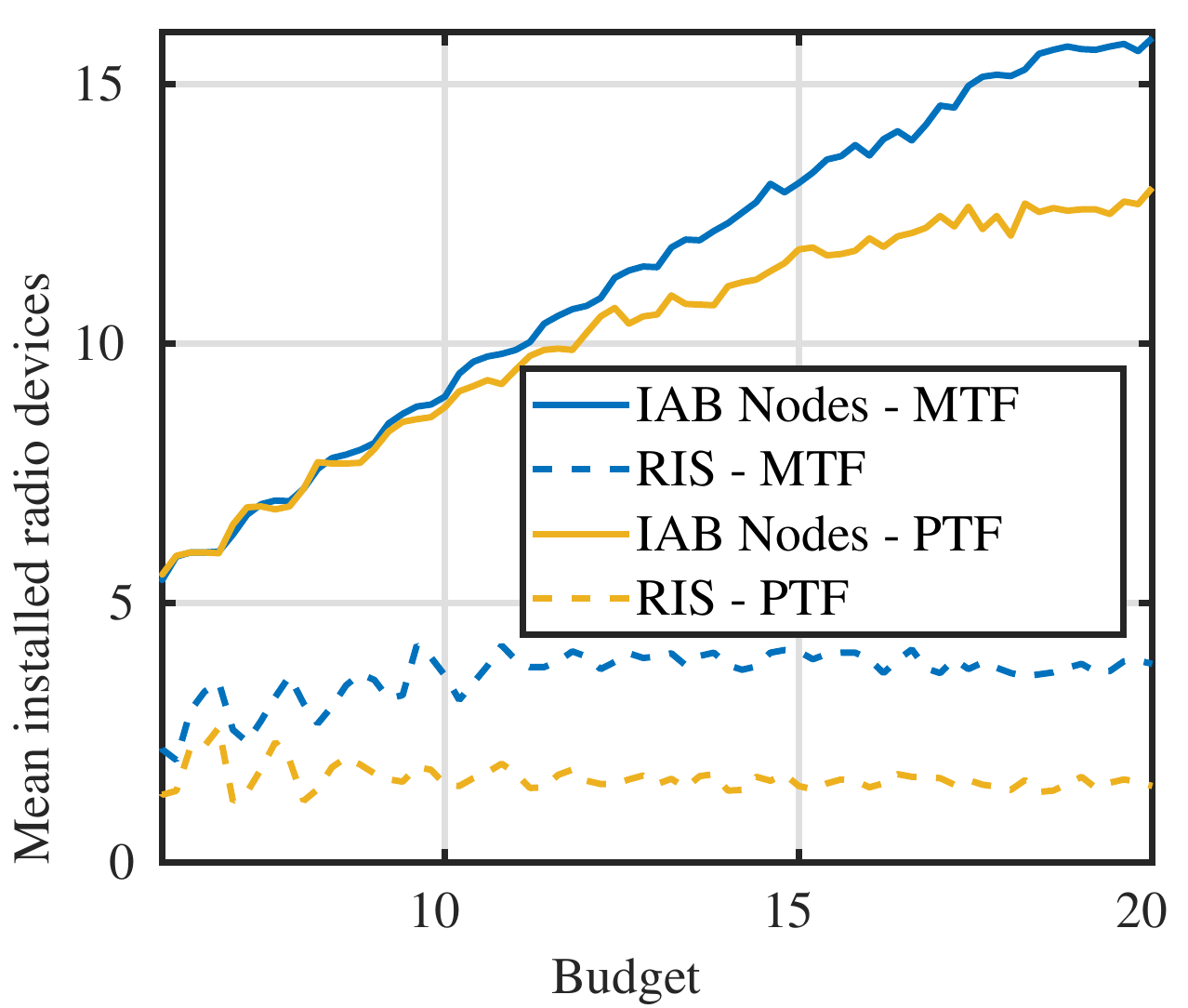}
\label{fig:devices}}
\hfil
\subfloat[Average throughputs per user]{\includegraphics[width=1.6in]{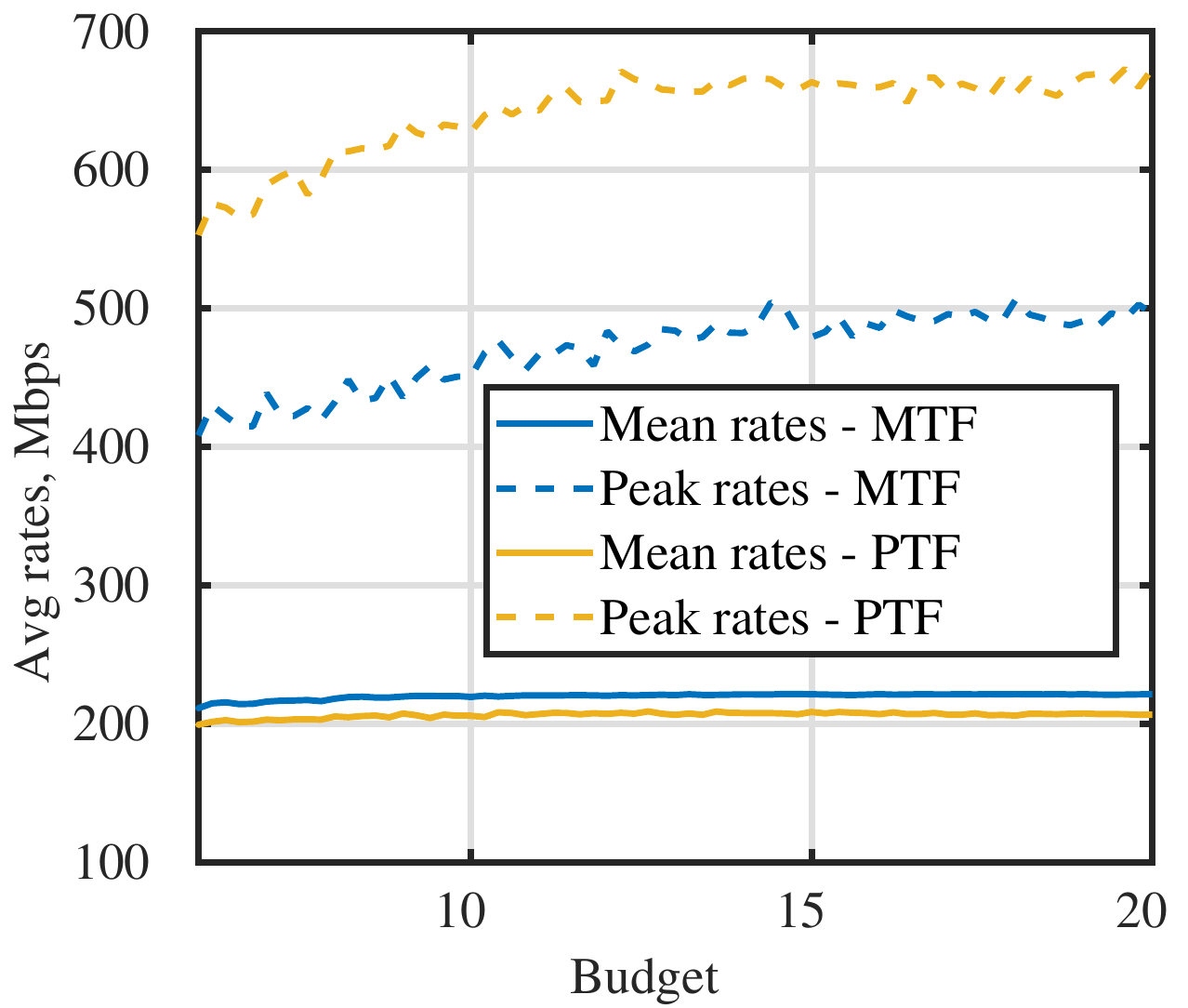}
\label{fig:rates}}
\caption{Device installation and throughput sensitivity to budget variation.}
\label{fig:performance}
\vspace{-5mm}
\end{figure}
\subsection{Performance Comparison}
Figure~\ref{fig:devices} shows how RIS usage is more prominent in the MTF planning (dashed blue curve), where, on average, four RIS are installed in each cell; in PTF (dashed yellow curve), instead, RIS reduces to about 2 per scenario, indicating that they are not suitable devices to increase peak throughput. As for IAB Nodes, the two models perform in the same way up to around 10 units of budget. Beyond that point, MTF (solid blue curve) tends to buy more IAB Nodes compared to PTF (solid yellow curve). In Figure~\ref{fig:rates}, the metrics driving the whole optimization are shown: the solid lines represent the mean throughput achievable in cells planned with MTF (blue) and PTF (yellow), while the dashed lines indicate the related peak throughput. It is clear from these results that even if PTF's results derive from a heuristic approach, peak throughputs in PTF-planned layouts gain more than $150 Mbps$ compared to those of the same scenarios planned with MTF, while at the same time obtaining a similar mean throughput. Remarkable is the fact that there is no need to increase the available budget significantly above 12 units since the mean throughput is almost constant throughout the plot, and the peak throughput does not show meaningful improvement, further increasing the budget.\\
These results show that the PTF-planned networks obtain a quasi-optimal mean throughput (i.e., like MTF planning) while notably improving the peak throughput for the users, so they can fully exploit the entire capacity of the RAN.

\subsection{Topology Features}
\begin{figure}[!t]
\centering
\subfloat[Average number of hops per user]{\includegraphics[width=1.6in]{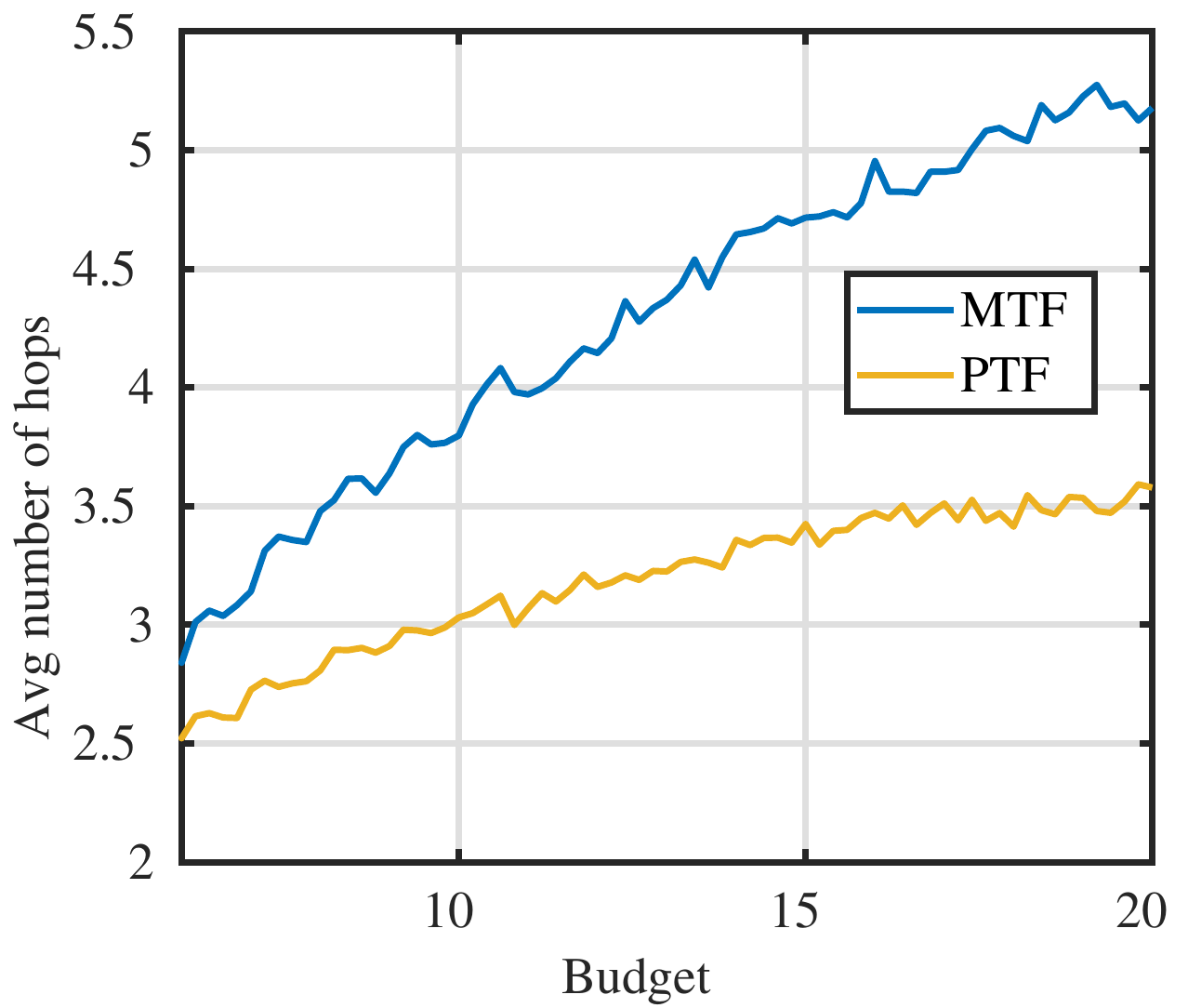}
\label{fig:multihops}}
\hfil
\subfloat[Average degree of IAB Donor]{\includegraphics[width=1.6in]{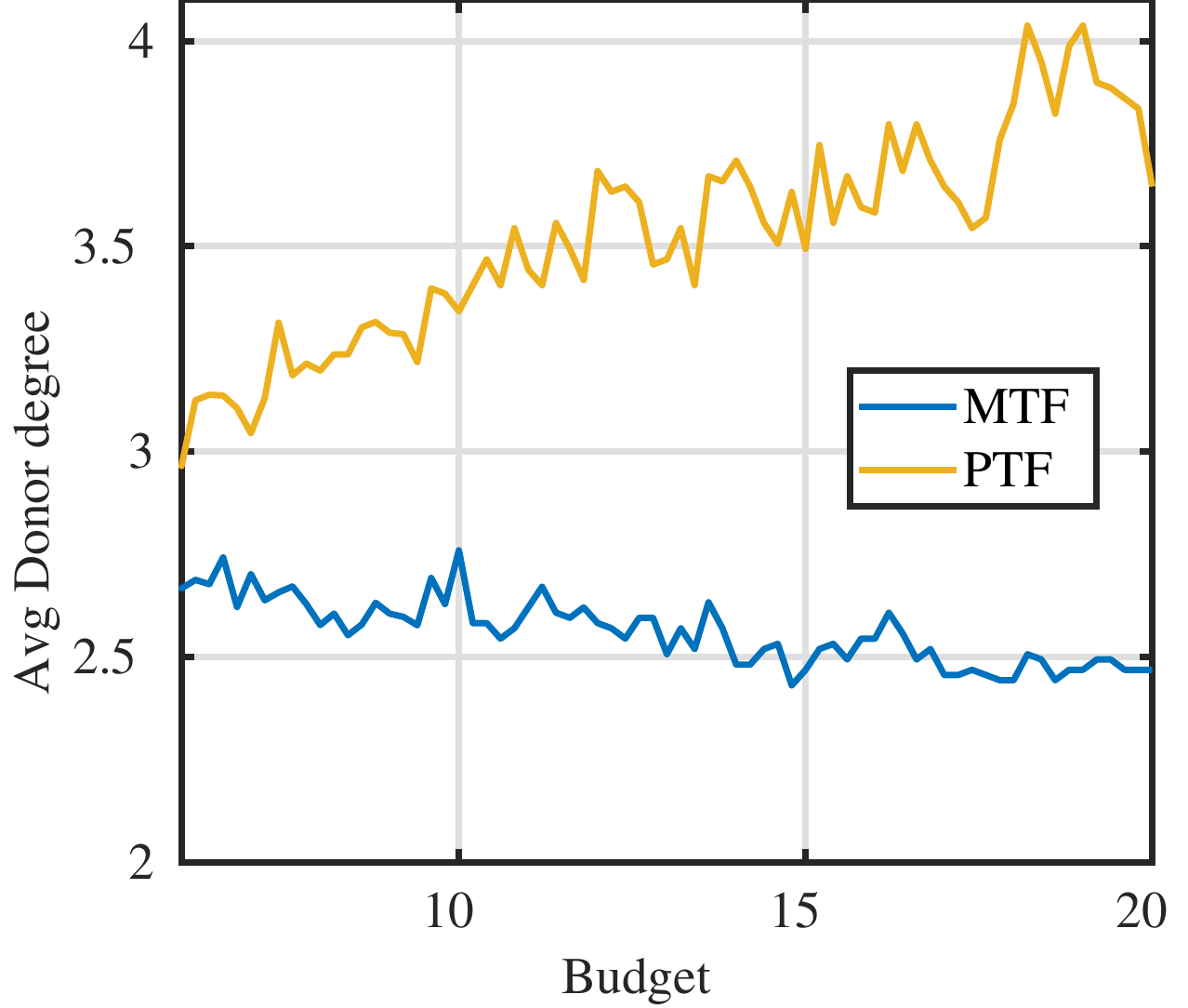}
\label{fig:donor_degree}}
\caption{Multi-hop and Donor degree sensitivity to budget variation.}
\label{fig:topology}
\vspace{-5mm}
\end{figure}
\begin{figure}[!t]
\centering
\subfloat[MTF topology example, top view.]{\includegraphics[width=3in]{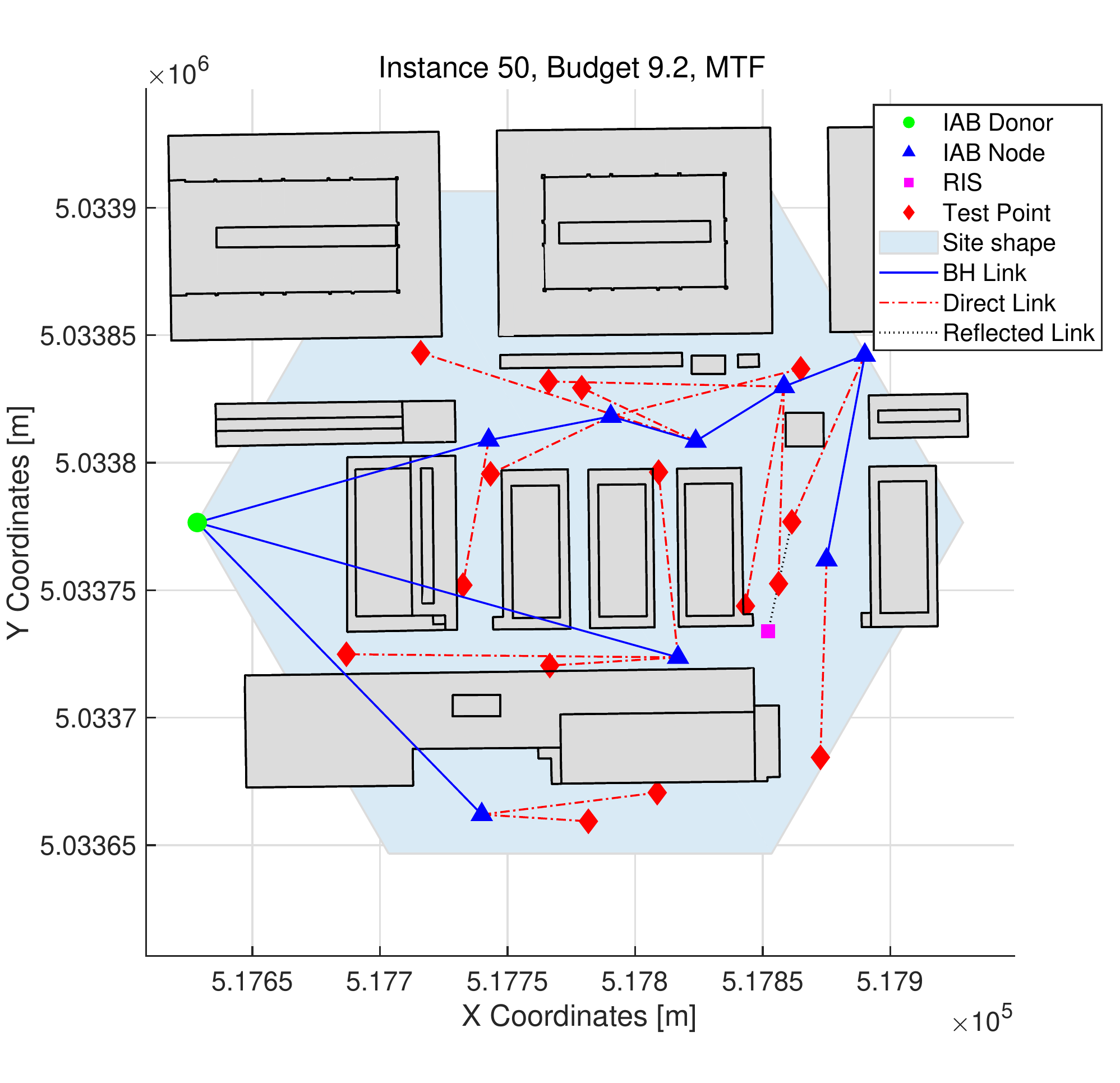}
\label{fig:MTF}}
\hfil
\subfloat[PTF topology example, top view.]{\includegraphics[width=3in]{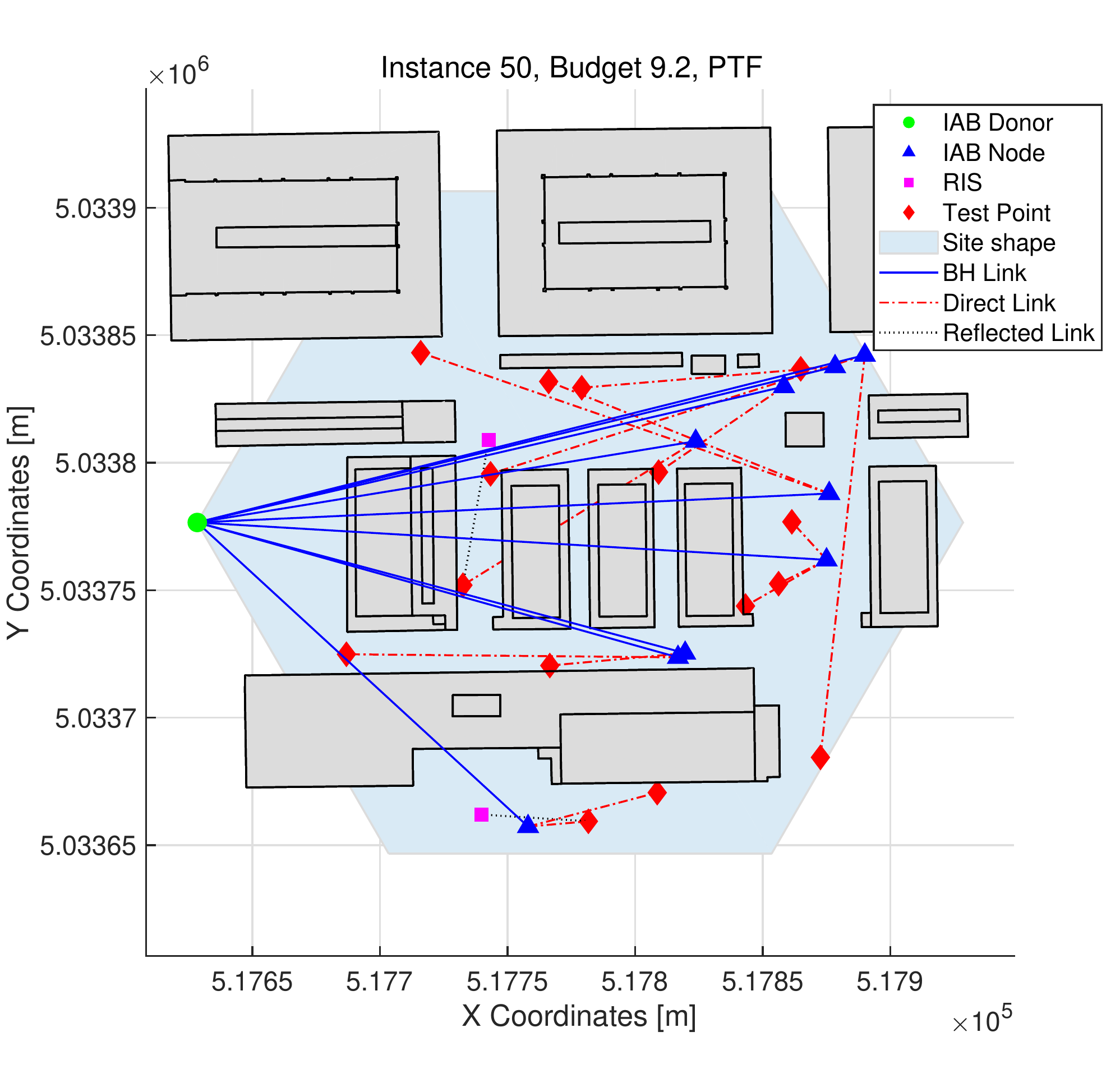}
\label{fig:PTF}}
\caption{Topology comparison of the two formulations from the top.}
\label{fig:comparison}
\vspace{-5mm}
\end{figure}
\begin{figure}[!t]
\centering
\subfloat[MTF topology example, isometric view.]{\includegraphics[width=247pt]{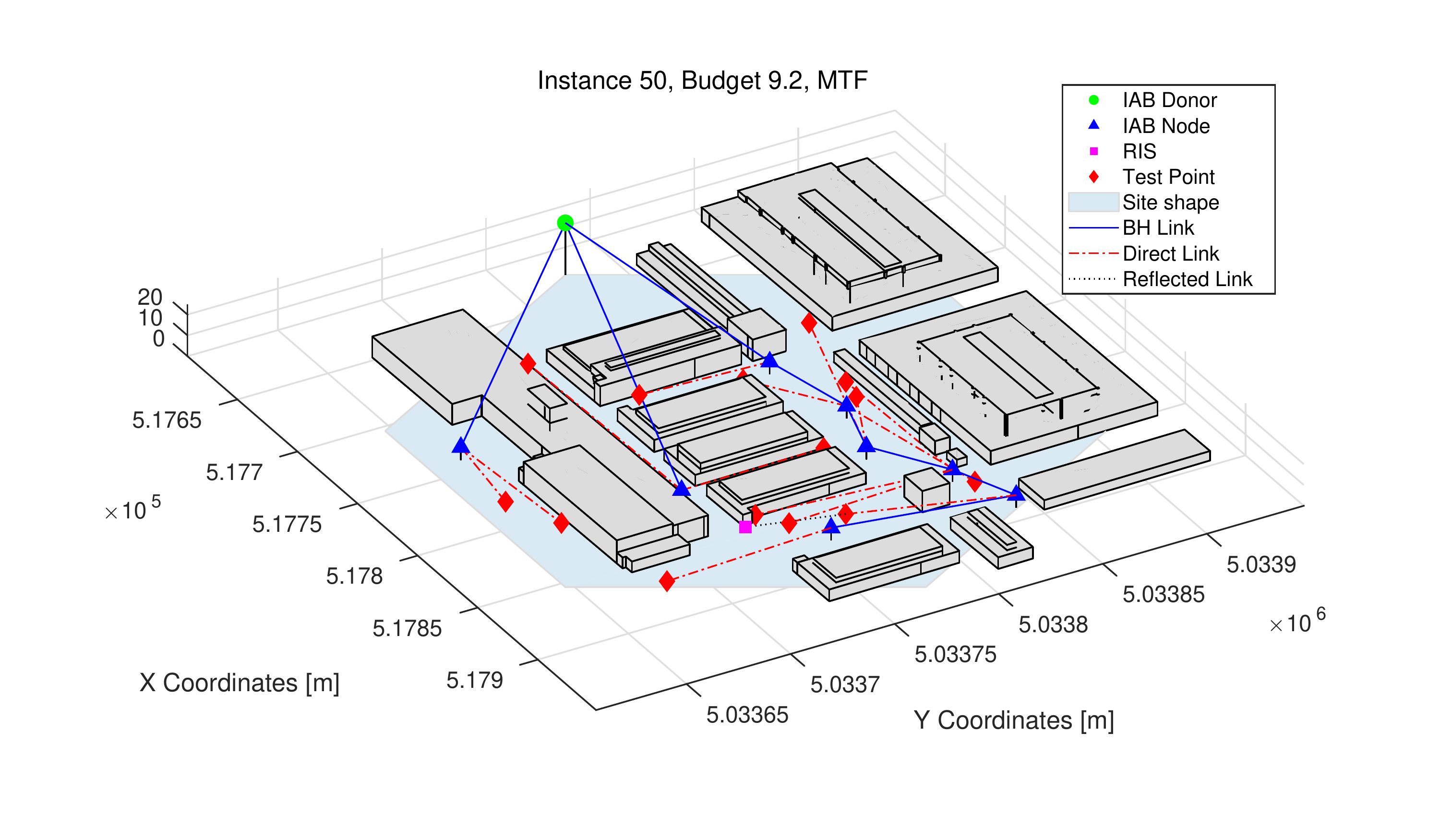}
\label{fig:MTF3d}}
\hfil
\subfloat[PTF topology example, isometric view.]{\includegraphics[width=247pt]{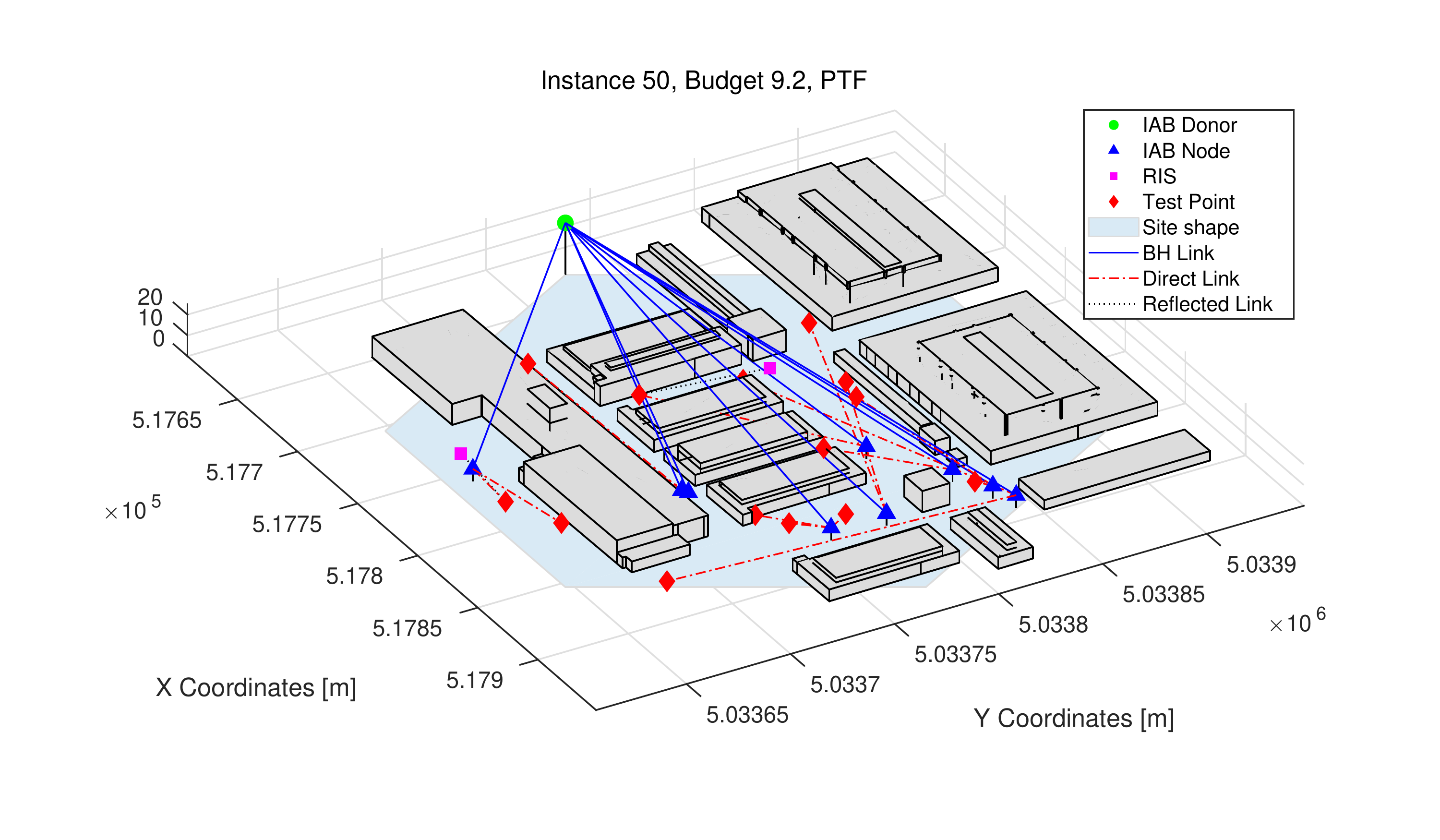}
\label{fig:PTF3d}}
\caption{Topology comparison of the two formulations from an isometric view.}
\label{fig:comparison3d}
\vspace{-5mm}
\end{figure}

In this subsection, we investigate how the different throughput performance reflects in a different network topology generated by the two formulations.
Since the IAB network is intrinsically connected to the concept of multi-hop forwarding, it is important to check the average path depth of the tree from the IAB Donor to a generic UE, measured in number of hops. Together with the depth, another relevant aspect of a tree topology is its degree, that is, how many subtrees spawn from each node. In particular, the degree of the root node (the IAB Donor) can be used as a metric to evaluate how close to a star topology the RAN is.
In Figure~\ref{fig:multihops}, PTF obtains a consistently smaller average number of hops per user than MTF, indicating that its trees must be shallower than those generated by MTF. The degree of the Donor is shown in Figure~\ref{fig:donor_degree}. PTF not only assigns the Donor more incident links than MTF, but the two curves have opposite trends as the available budget increases: PTF gets closer to a star topology, while MTF departs from it. As a proof of concept, we report in Figure~\ref{fig:comparison} an instance with a significant difference between the peak throughput obtained by the two models; it is evident how PTF (Figure~\ref{fig:PTF}) selects star-like topologies compared to MTF (Figure~\ref{fig:MTF}). For completeness, we include the same instance, shown from a three-dimensional isometric view in Figure~\ref{fig:comparison3d}.

\vspace{-2mm}
\section{Conclusion}
\label{sec:conclusion}
In this paper, we analyzed the potential of mm-Wave IAB RANs equipped with RISs to provide peak and mean throughput traffic.
Our analysis has demonstrated that optimizing the network layout of mm-Wave RANs integrated with IAB and RIS considering the peak user throughput can significantly improve the achievable peak throughput in realistic urban scenarios compared to the traditional mean-throughput approach. In addition, these layouts can guarantee a mean throughput comparable with the one achievable via traditional planning approaches.\\
Network layouts must move towards shallow star-like topology to pursue peak throughput maximization. This has important practical implications for developing and deploying next-generation networks, particularly in urban environments, where the demand for ultra-high-speed connectivity is rapidly increasing. 
\vspace{-3mm}
\section*{Acknowledgment}
The research in this paper has been carried out in the framework of
Huawei-Politecnico di Milano Joint Research Lab. The Authors
acknowledge Huawei Milan research center for the
collaboration.

\vspace{-2mm}
\bibliographystyle{IEEEtran}
\bibliography{biblio.bib}

\begin{thebibliography}{10}
\providecommand{\url}[1]{#1}
\csname url@samestyle\endcsname
\providecommand{\newblock}{\relax}
\providecommand{\bibinfo}[2]{#2}
\providecommand{\BIBentrySTDinterwordspacing}{\spaceskip=0pt\relax}
\providecommand{\BIBentryALTinterwordstretchfactor}{4}
\providecommand{\BIBentryALTinterwordspacing}{\spaceskip=\fontdimen2\font plus
\BIBentryALTinterwordstretchfactor\fontdimen3\font minus
  \fontdimen4\font\relax}
\providecommand{\BIBforeignlanguage}[2]{{%
\expandafter\ifx\csname l@#1\endcsname\relax
\typeout{** WARNING: IEEEtran.bst: No hyphenation pattern has been}%
\typeout{** loaded for the language `#1'. Using the pattern for}%
\typeout{** the default language instead.}%
\else
\language=\csname l@#1\endcsname
\fi
#2}}
\providecommand{\BIBdecl}{\relax}
\BIBdecl

\bibitem{rapparport2014}
S.~{Rangan}, T.~S. {Rappaport}, and E.~{Erkip}, ``Millimeter-wave cellular
  wireless networks: Potentials and challenges,'' \emph{Proceedings of the
  IEEE}, vol. 102, no.~3, pp. 366--385, 2014.

\bibitem{Akdeniz2014}
M.~R. Akdeniz, Y.~Liu, M.~K. Samimi, S.~Sun, S.~Rangan, T.~S. Rappaport, and
  E.~Erkip, ``{Millimeter wave channel modeling and cellular capacity
  evaluation},'' \emph{IEEE Jrn. on Sel. Areas in Comm.}, vol.~32, no.~6, pp.
  1164--1179, 2014.

\bibitem{ericssonReportNov2022}
\BIBentryALTinterwordspacing
Ericsson. (2022) "ericsson mobility report: November 2022". [Online].
  Available:
  \url{https://www.ericsson.com/en/reports-and-papers/mobility-report/reports/november-2022}
\BIBentrySTDinterwordspacing

\bibitem{8254900}
G.~R. MacCartney, T.~S. Rappaport, and S.~Rangan, ``Rapid fading due to human
  blockage in pedestrian crowds at 5g millimeter-wave frequencies,'' in
  \emph{GLOBECOM 2017 - 2017 IEEE Global Communications Conference}, 2017, pp.
  1--7.

\bibitem{10.3389/frcmn.2021.647284}
\BIBentryALTinterwordspacing
Y.~Zhang, M.~A. Kishk, and M.-S. Alouini, ``A survey on integrated access and
  backhaul networks,'' \emph{Frontiers in Communications and Networks}, vol.~2,
  2021. [Online]. Available:
  \url{https://www.frontiersin.org/articles/10.3389/frcmn.2021.647284}
\BIBentrySTDinterwordspacing

\bibitem{938713}
H.~Willebrand and B.~Ghuman, ``Fiber optics without fiber,'' \emph{IEEE
  Spectrum}, vol.~38, no.~8, pp. 40--45, 2001.

\bibitem{8514996}
M.~Polese, M.~Giordani, A.~Roy, S.~Goyal, D.~Castor, and M.~Zorzi, ``End-to-end
  simulation of integrated access and backhaul at mmwaves,'' in \emph{2018 IEEE
  23rd International Workshop on Computer Aided Modeling and Design of
  Communication Links and Networks (CAMAD)}, 2018, pp. 1--7.

\bibitem{9040265}
M.~Polese, M.~Giordani, T.~Zugno, A.~Roy, S.~Goyal, D.~Castor, and M.~Zorzi,
  ``Integrated access and backhaul in 5g mmwave networks: Potential and
  challenges,'' \emph{IEEE Communications Magazine}, vol.~58, no.~3, pp.
  62--68, 2020.

\bibitem{9720231}
R.~Flamini, D.~De~Donno, J.~Gambini, F.~Giuppi, C.~Mazzucco, A.~Milani, and
  L.~Resteghini, ``Toward a heterogeneous smart electromagnetic environment for
  millimeter-wave communications: An industrial viewpoint,'' \emph{IEEE
  Transactions on Antennas and Propagation}, vol.~70, no.~10, pp. 8898--8910,
  2022.

\bibitem{Moro2021}
E.~Moro, I.~Filippini, A.~Capone, and D.~{De Donno}, ``Planning {Mm-Wave}
  access networks with reconfigurable intelligent surfaces,'' in \emph{2021
  IEEE 32nd Annual Int. Sym. on Personal, Indoor and Mobile Radio Comm.
  (PIMRC)}, Helsinki, Finland, Sep. 2021.

\bibitem{9771934}
P.~Fiore, E.~Moro, I.~Filippini, A.~Capone, and D.~D. Donno, ``Boosting 5g
  mm-wave iab reliability with reconfigurable intelligent surfaces,'' in
  \emph{2022 IEEE Wireless Communications and Networking Conference (WCNC)},
  2022, pp. 758--763.

\bibitem{9930587}
G.~Leone, E.~Moro, I.~Filippini, A.~Capone, and D.~D. Donno, ``Towards reliable
  mmwave 6g ran: Reconfigurable surfaces, smart repeaters, or both?'' in
  \emph{2022 20th International Symposium on Modeling and Optimization in
  Mobile, Ad hoc, and Wireless Networks (WiOpt)}, 2022, pp. 81--88.

\bibitem{tsilipakos2020toward}
O.~Tsilipakos, A.~C. Tasolamprou, A.~Pitilakis, F.~Liu, X.~Wang, M.~S.
  Mirmoosa, D.~C. Tzarouchis, S.~Abadal, H.~Taghvaee, C.~Liaskos \emph{et~al.},
  ``Toward intelligent metasurfaces: The progress from globally tunable
  metasurfaces to software-defined metasurfaces with an embedded network of
  controllers,'' \emph{Advanced Optical Materials}, vol.~8, no.~17, p. 2000783,
  2020.

\bibitem{di2019smart}
M.~Di~Renzo, M.~Debbah, D.-T. Phan-Huy, A.~Zappone, M.-S. Alouini, C.~Yuen,
  V.~Sciancalepore, G.~C. Alexandropoulos, J.~Hoydis, H.~Gacanin \emph{et~al.},
  ``Smart radio environments empowered by reconfigurable ai meta-surfaces: An
  idea whose time has come,'' \emph{EURASIP Journal on Wireless Communications
  and Networking}, vol. 2019, no.~1, pp. 1--20, 2019.

\bibitem{6678102}
H.~Ju and R.~Zhang, ``Throughput maximization in wireless powered communication
  networks,'' \emph{IEEE Transactions on Wireless Communications}, vol.~13,
  no.~1, pp. 418--428, 2014.

\bibitem{sandvineReport2022}
\BIBentryALTinterwordspacing
Sandvine. (2022) "2022 global internet phenomena report". [Online]. Available:
  \url{https://www.sandvine.com/global-internet-phenomena-report-2022}
\BIBentrySTDinterwordspacing

\bibitem{sandvineReport2023}
\BIBentryALTinterwordspacing
------. (2023) "2023 global internet phenomena report". [Online]. Available:
  \url{https://www.sandvine.com/global-internet-phenomena-report-2023}
\BIBentrySTDinterwordspacing

\bibitem{3GPPTR38.874}
3GPP, ``3rd generation partnership project; technical specification group radio
  access network; nr; study on integrated access and backhaul; (release 16),''
  3GPP, Tech. Rep., 2018.

\bibitem{devoti2020}
F.~{Devoti} and I.~{Filippini}, ``Planning mm-wave access networks under
  obstacle blockages: A reliability-aware approach,'' \emph{IEEE/ACM Trans. on
  Networking}, vol.~28, no.~5, pp. 2203--2214, 2020.

\bibitem{7342886}
S.~Kutty and D.~Sen, ``Beamforming for millimeter wave communications: An
  inclusive survey,'' \emph{IEEE Communications Surveys \& Tutorials}, vol.~18,
  no.~2, pp. 949--973, 2016.

\bibitem{https://doi.org/10.48550/arxiv.2211.08033}
\BIBentryALTinterwordspacing
R.~A. Ayoubi, M.~Mizmizi, D.~Tagliaferri, D.~De~Donno, and U.~Spagnolini,
  ``Network-controlled repeaters vs. reconfigurable intelligent surfaces for 6g
  mmw coverage extension,'' 2022. [Online]. Available:
  \url{https://arxiv.org/abs/2211.08033}
\BIBentrySTDinterwordspacing

\bibitem{GRANW2018study}
G.~R. A. N.~W. Group, ``Study on channel model for frequencies from 0.5 to 100
  ghz (release 15),'' 3GPP, Tech. Rep., 2020.

\bibitem{doi:https://doi.org/10.1002/9781118511305.ch7}
\BIBentryALTinterwordspacing
A.~Capone, I.~Filippini, S.~Gualandi, and D.~Yuan, \emph{Resource Optimization
  in Multiradio Multichannel Wireless Mesh Networks}.\hskip 1em plus 0.5em
  minus 0.4em\relax John Wiley \& Sons, Ltd, 2013, ch.~7, pp. 239--274.
  [Online]. Available:
  \url{https://onlinelibrary.wiley.com/doi/abs/10.1002/9781118511305.ch7}
\BIBentrySTDinterwordspacing

\bibitem{MilanoGeoportale}
\BIBentryALTinterwordspacing
(2012) "milano geoportale - open data". Comune di Milano. [Online]. Available:
  \url{https://geoportale.comune.milano.it/sit/open-data/}
\BIBentrySTDinterwordspacing

\end{thebibliography}

\clearpage
\newpage

\end{document}